\documentclass[10pt,a4paper]{article}

\usepackage{titlesec}
\titleformat*{\section}{\bf\normalsize\large}
\titleformat*{\subsection}{\bf\normalsize}

\titlespacing{\section}{0pt}{2ex}{1ex}
\titlespacing{\subsection}{0pt}{2ex}{1ex}
\titlespacing{\subsubsection}{0pt}{2ex}{1ex}

\usepackage{fancyhdr}
\usepackage[latin1]{inputenc}
\usepackage[top=2cm,bottom=2cm,left=2cm,right=2cm]{geometry}
\usepackage[usenames,dvipsnames]{color}
\usepackage[pdfborder={0 0 0},colorlinks=true,urlcolor=blue,linkcolor=blue,citecolor=blue,bookmarks=true]{hyperref}
\hypersetup{pdftitle={PyArmadillo: a streamlined linear algebra library for Python}}
\hypersetup{pdfauthor={Jason Rumengan, Terry Yue Zhuo, Conrad Sanderson}}
\usepackage{graphicx}
\usepackage{multirow}
\usepackage[british]{babel}
\usepackage[none]{hyphenat}
\usepackage{caption}
\usepackage{palatino}

\usepackage[newfloat,frozencache]{minted}  % use the cache; pdflatex paper

\makeatletter
\renewcommand\@makefntext[1]{%
\noindent
\mbox{\@thefnmark}{#1}}
\makeatother

\newcommand\blfootnote[1]{%
  \begingroup
  \renewcommand\thefootnote{}\footnote{#1}%
  \addtocounter{footnote}{-1}%
  \endgroup
}

\def\us{{\_}}

\begin{document}

\renewcommand{\baselinestretch}{1.05}\small\normalsize

\thispagestyle{empty}
\pagestyle{empty}

\begin{minipage}{1\textwidth}
\centering
\mbox{\LARGE\bf PyArmadillo: a streamlined linear algebra library for Python}\\
\vspace{3ex}

{\large Jason Rumengan~\textsuperscript{$\dagger\ddagger$}, Terry Yue Zhuo~\textsuperscript{$\diamond$}, Conrad Sanderson~\textsuperscript{$\dagger\ast$}}\\
\vspace{2ex}

\begin{minipage}{0.5\textwidth}
\small
\textsuperscript{$\dagger$} Data61/CSIRO, Australia\\
\textsuperscript{$\ddagger$} Queensland University of Technology, Australia\\
\textsuperscript{$\diamond$} University of New South Wales, Australia\\
\textsuperscript{$\ast$} Griffith University, Australia
\end{minipage}
\end{minipage}

\vspace{3ex}

\section*{Abstract}

PyArmadillo is a linear algebra library for the Python language,
%PyArmadillo is a linear algebra and scientific computing library for the Python language,
with the aim of closely mirroring the programming interface of the widely used Armadillo C++ library, 
which in turn is deliberately similar to Matlab.
PyArmadillo hence facilitates algorithm prototyping with Matlab-like syntax directly in Python,
and relatively straightforward conversion of PyArmadillo-based Python code
into performant Armadillo-based C++ code.
The converted code can be used for purposes such as speeding up Python-based programs
in conjunction with {\it pybind11},
or the integration of algorithms originally prototyped in Python into larger C++ codebases.
PyArmadillo provides objects for matrices and cubes,
as well as over 200 associated functions for manipulating data stored in the objects.
Integer, floating point and complex numbers are supported.
Various matrix factorisations are provided through integration with LAPACK,
or one of its high performance drop-in replacements such as Intel~MKL or OpenBLAS.
PyArmadillo is open-source software, distributed under the Apache 2.0 license;
it can be obtained at
\href{https://pyarma.sourceforge.io}{https://pyarma.sourceforge.io}
or via the Python Package Index in precompiled form.
\blfootnote{\textbf{Published in:} Journal of Open Source Software, Vol.~6, No.~66, 2021. \textbf{DOI:} \href{https://doi.org/10.21105/joss.03051}{10.21105/joss.03051}}

\vspace{2ex}

\section*{Background}

Armadillo is a popular linear algebra and scientific computing library for the C++ language~\cite{Sanderson_2016,Sanderson_2018}
that has three main characteristics:
{\bf (i)}~a high-level programming interface deliberately similar to Matlab,
{\bf (ii)}~an expression evaluator (based on template meta-programming)
that automatically combines several operations to increase speed and efficiency,
and
{\bf (iii)}~an efficient mapper between mathematical expressions and low-level BLAS/LAPACK functions~\cite{Psarras_2021,Sanderson_2020}.
Matlab is widely used in both industrial and academic contexts,
providing a programming interface that allows mathematical expressions to be written in a concise and natural manner~\cite{Linge_MatlabOctave_2016}, 
especially in comparison to directly using low-level libraries such as LAPACK~\cite{anderson1999lapack}.
In industrial settings, algorithms are often first prototyped in Matlab,
before conversion into another language, such as C++, for the purpose of integration into products.
The similarity of the programming interfaces between Armadillo and Matlab
facilitates direct prototyping in C++,
as well as the conversion of research code into production environments.
Armadillo is also often used for implementing performance critical parts of software packages
running under the R environment for statistical computing~\cite{R_manual}, 
via the RcppArmadillo bridge~\cite{Eddelbuettel_2014}.

Over the past few years, Python has become popular for data science and machine learning.
This partly stems from a rich ecosystem of supporting frameworks and packages,
as well as lack of licensing costs in comparison to Matlab.
Python allows relatively quick prototyping of algorithms,
aided by its dynamically typed nature
and the interpreted execution of user code,
avoiding time-consuming compilation into machine code.
However, for the joint purpose of algorithm prototyping and deployment,
the flexibility of Python comes with two main issues:
{\bf (i)}~slow execution speed due to the interpreted nature of the language,
{\bf (ii)}~difficulty with integration of code written in Python into larger programs and/or frameworks written in another language.
The first issue can be somewhat addressed through conversion of Python-based code into the low-level Cython language~\cite{Cython}.
However, since Cython is closely tied with Python, 
conversion of Python code into C++ may be preferred as it also addresses the second issue,
as well as providing a higher-level of abstraction.

\newpage

PyArmadillo is aimed at:
{\bf (i)}~users that prefer compact Matlab-like syntax rather than the somewhat more verbose syntax provided by NumPy/SciPy~\cite{harris2020array,2020SciPy-NMeth},
and
{\bf (ii)}~users that would like a straightforward conversion path to performant C++ code.
More specifically,
PyArmadillo aims to closely mirror the programming interface of the Armadillo library,
thereby facilitating the prototyping of algorithms with Matlab-like syntax directly in Python.
Furthermore, PyArmadillo-based Python code can be easily converted
into high-performance Armadillo-based C++ code.
Due to the similarity of the programming interfaces,
the risk of introducing bugs in the conversion process is considerably reduced.
Moreover, conversion into C++ based code allows taking advantage of expression optimisation
performed at compile-time by Armadillo, resulting in further speedups.
The resulting code can be used in larger C++ programs, 
or used as a replacement of performance critical parts within a Python program
with the aid of the {\it pybind11} interface layer~\cite{pybind11}.

\section*{Functionality}

PyArmadillo provides matrix objects for several distinct element types: integers, single- and double-precision floating point numbers, as well as complex numbers.
In addition to matrices, PyArmadillo also has support for cubes (3 dimensional arrays), where each cube can be treated as an ordered set of matrices.
% Multi-dimensional arrays beyond 3 dimensions are explicitly beyond the scope of PyArmadillo.
Over 200 functions are provided for manipulating data stored in the objects, covering the following areas:
\linebreak
fundamental arithmetic operations, 
contiguous and non-contiguous submatrix~views,
diagonal~views, 
\linebreak
element-wise functions,
scalar/vector/matrix valued functions of matrices,
% generation~of~various vectors/matrices,
statistics,
signal processing,
storage of matrices in files,
matrix decompositions/factorisations,
matrix inverses,
and equation solvers.
PyArmadillo matrices and cubes are convertible to/from NumPy arrays,
allowing users to tap into the wider Python data science ecosystem,
including plotting tools such as {\it Matplotlib}~\cite{Hunter2007}.

An overview of the available functionality in PyArmadillo (as of version 0.500) is given in Tables~\ref{tab:mat_class_members} through to~\ref{tab:signal_processing_functions}.
\linebreak
Table~\ref{tab:mat_class_members} briefly describes the member functions and variables of the main matrix class;
Table~\ref{tab:operators} lists the main subset of overloaded Python operators;
Table~\ref{tab:matrix_decompositions} outlines matrix decompositions and equation solvers;
\linebreak
Table~\ref{tab:matrix_generators} overviews functions for generating vectors with various sequences;
Table~\ref{tab:functions_of_matrices} lists the main forms of general functions of matrices;
Table~\ref{tab:elementwise_functions} lists various element-wise functions of matrices;
Table~\ref{tab:statistics_functions} summarises the set of functions and classes focused on statistics;
Table~\ref{tab:signal_processing_functions} lists functions for signal and image processing.
\linebreak
Lastly, 
Table~\ref{tab:matlab_armadillo_syntax} 
provides examples of Matlab syntax and conceptually corresponding PyArmadillo syntax.
Figure~\ref{fig:simple_prog} shows a simple PyArmadillo based Python program.
See the online documentation at \href{https://pyarma.sourceforge.io/docs.html}{https://pyarma.sourceforge.io/docs.html}
for more details and examples.

\begin{figure}[!h]
\centering
\vspace{1ex}
\begin{minipage}{0.4\textwidth}
\hrule
\vspace{1ex}
% \begin{verbatim}
\begin{minted}[fontsize=\footnotesize,escapeinside=||]{python}
from pyarma import *

A = |\textcolor{blue}{mat}|(4, 5, |\textcolor{red}{fill.ones}|)
B = |\textcolor{blue}{mat}|(4, 5, |\textcolor{red}{fill.randu}|)

C = A * B.t()

C.print("C:")
\end{minted}
% \end{verbatim}
\hrule
\end{minipage}
\caption
  {
  A short PyArmadillo based Python program.
  See Tables~\ref{tab:mat_class_members} to~\ref{tab:signal_processing_functions}
  for an overview of the available functionality.}
\label{fig:simple_prog}
\end{figure}

\section*{Implementation \& License}

PyArmadillo relies on pybind11~\cite{pybind11} for interfacing C++ and Python,
as well as on Armadillo~\cite{Sanderson_2016,Sanderson_2018} for the underlying C++ implementation of matrix objects and associated functions.
Due to its expressiveness and relatively straightforward use,
pybind11 was selected over other interfacing approaches such as Boost.Python~\cite{osti_815409} and manually writing C++ extensions for Python.
In turn, Armadillo interfaces with low-level routines in BLAS and LAPACK~\cite{anderson1999lapack},
where BLAS is used for matrix multiplication,
and LAPACK is used for various matrix decompositions/factorisations and equation solvers.
As the low-level routines in BLAS and LAPACK are considered as a {\it de facto} standard for numerical linear algebra,
it is possible to use high performance drop-in replacements such as Intel~MKL~\cite{Intel_MKL} and OpenBLAS~\cite{Xianyi_OpenBLAS}.
PyArmadillo is open-source software available under the permissive Apache License 2.0~\cite{apache},
making it useful in both open-source and proprietary (closed-source) contexts~\cite{Laurent_2008}.

\newpage

\begin{table}[!htb]
\centering
\small
\caption
  {
  Subset of member functions and variables of the {\it mat} class, the main matrix object in PyArmadillo.
  }
\label{tab:mat_class_members}
\vspace{-1ex}
\begin{tabular}{l|lllllllll}
\hline
{\bf Function/Variable}  & {\bf Description} \\
\hline
$.$n{\us}rows           & number of rows (read only)\\
$.$n{\us}cols           & number of columns (read only)\\
$.$n{\us}elem           & total number of elements (read only)\\
\hline
$[$i$]$                & access the {\it i}-th element, assuming a column-by-column layout\\
$[$r,~c$]$             & access the element at row {\it r} and column {\it c}\\
\hline
$.$in{\us}range(i)     & test whether the {\it i}-th element can be accessed\\
$.$in{\us}range(r,~c)  & test whether the element at row {\it r} and column {\it c} can be accessed\\
\hline
$.$reset()           & set the number of elements to zero\\
$.$copy{\us}size(A)     & set the size to be the same as matrix {\it A}\\
$.$set{\us}size(n{\us}rows, n{\us}cols) & change size to specified dimensions, without preserving data (fast)\\
.reshape(n{\us}rows, n{\us}cols)    & change size to specified dimensions, with elements copied column-wise (slow)\\
.resize(n{\us}rows, n{\us}cols)    & change size to specified dimensions, while preserving elements \& their layout (slow)\\
\hline
$.$ones(n{\us}rows, n{\us}cols)    & set all elements to one, optionally first resizing to specified dimensions\\
$.$zeros(n{\us}rows, n{\us}cols)   & as above, but set all elements to zero\\
$.$randu(n{\us}rows, n{\us}cols)   & as above, but set elements to uniformly distributed random values in [0,1] interval\\
$.$randn(n{\us}rows, n{\us}cols)   & as above, but use a Gaussian/normal distribution with $\mu$ = 0 and $\sigma$ = 1\\
$.$fill(k)            & set all elements to be equal to {\it k}\\
$.$for{\us}each(lambda val : {\footnotesize ...}) & for each element, pass its value to a lambda function\\
\hline
$.$is{\us}empty()      & test whether there are no elements\\
$.$is{\us}finite()      & test whether all elements are finite\\
$.$is{\us}square()      & test whether the matrix is square\\
$.$is{\us}vec()         & test whether the matrix is a vector \\
$.$is{\us}sorted()      & test whether the matrix is sorted \\
\hline
$.$has{\us}inf()      & test whether any element is $\pm\infty$\\
$.$has{\us}nan()      & test whether any element is not-a-number\\
\hline
iter(A)               & iterator of matrix {\it A}\\
iterator(A, i, j)     & iterator of matrix {\it A} from the {\it i}-th element to the {\it j}-th\\
col{\us}iter(A, i, j)   & iterator of matrix {\it A} from the {\it i}-th column to the {\it j}-th\\
row{\us}iter(A, i, j)   & iterator of matrix {\it A} from the {\it i}-th row to the {\it j}-th\\
\hline
$.$print(header)     & print elements, with an optional text header\\
\hline
$.$save(name, format)      & store matrix in the specified file, optionally specifying storage format\\
$.$load(name, format)      & retrieve matrix from the specified file, optionally specifying format\\
\hline
$[$diag, k$]$         & read/write access to {\it k}-th diagonal\\
$[$i, :$]$          & read/write access to row {\it i}\\
$[$:, i$]$          & read/write access to column {\it i}\\
$[$a:b, :$]$      & read/write access to submatrix, spanning from row {\it a} to row {\it b} \\
$[$:, c:d$]$      & read/write access to submatrix, spanning from column {\it c} to column {\it d} \\
$[$a:b, c:d$]$ & read/write access to submatrix spanning rows {\it a} to {\it b} and columns {\it c} to {\it d}\\
$[$p, q, size(A)$]$ & read/write access to submatrix starting at row {\it p} and col {\it q} with size same as matrix {\it A}\\
$[$vector{\us}of{\us}row{\us}indices, :$]$ & read/write access to rows corresponding to the specified indices \\
$[$:, vector{\us}of{\us}col{\us}indices$]$ & read/write access to columns corresponding to the specified indices \\
$[$vector{\us}of{\us}indices$]$ & read/write access to matrix elements corresponding to the specified indices\\
\hline
$.$swap{\us}rows(p, q)   & swap the contents of specified rows\\
$.$swap{\us}cols(p, q)   & swap the contents of specified columns\\
$.$insert{\us}rows(row, X) & insert a copy of {\it X} at the specified row \\
$.$insert{\us}cols(col, X) & insert a copy of {\it X} at the specified column \\
$.$shed{\us}rows(first{\us}row, last{\us}row) & remove the specified range of rows  \\
$.$shed{\us}cols(first{\us}col, last{\us}col) & remove the specified range of columns \\
\hline
$.$min()      & return minimum value\\
$.$max()      & return maximum value\\
$.$index{\us}min()      & return index of minimum value\\
$.$index{\us}max()      & return index of maximum value\\
\hline
\end{tabular}
\end{table}
\newpage

% operators:         & \\
% ~~$+$& \\
% ~~$-$& \\
% ~~$/$& \\
% ~~$*$& \\
% ~~$\%$& \\
% ~~$==$& \\
% ~~$!=$& \\
% ~~$>=$& \\
% ~~$<=$& \\
% ~~$>$& \\
% ~~$<$& \\

\begin{table}[!htb]
\centering
\small
\caption
  {
  Subset of matrix operations involving overloaded Python operators.
  %Caveat: to check for equality, it's more appropriate to use the {\it approx{\us}equal()} function,
  %due to the necessarily limited precision of the computational representation of numbers.
  }
\label{tab:operators}
\vspace{-1ex}
\begin{tabular}{l|lllllllll}
\hline
{\bf Operation}  & {\bf Description} \\
\hline
A $-$ k                  & subtract scalar {\it k} from all elements in matrix {\it A}\\
k $-$ A                  & subtract each element in matrix {\it A} from scalar {\it k} \\
A $+$ k, ~~ k $+$ A      & add scalar {\it k} to all elements in matrix {\it A}\\
A \hspace{0.3ex}$*$\hspace{0.3ex} k, ~~ k \hspace{0.3ex}$*$\hspace{0.3ex} A     & multiply matrix {\it A} by scalar {\it k} \\
\hline
A $+$ B                  & add matrices {\it A} and {\it B}\\
A $-$ B                  & subtract matrix {\it B} from {\it A}\\
A $*~$ B                 & matrix multiplication of {\it A} and {\it B}\\
A {\footnotesize $@$} B & element-wise multiplication of matrices {\it A} and {\it B}\\
A {\footnotesize $/$} ~B & element-wise division of matrix {\it A} by matrix {\it B}\\
\hline
A $==$ B                   & element-wise equality evaluation between matrices {\it A} and {\it B}\\
                           & [{\bf caveat}: use {\it approx{\us}equal()} to test whether all corresponding elements are approximately equal]\\
A ~~!$=$ B                   & element-wise non-equality evaluation between matrices {\it A} and {\it B}\\
\hline
A $>=$ B                   & element-wise evaluation whether elements in matrix {\it A} are greater-than-or-equal to elements in {\it B}\\
A $<=$ B                   & element-wise evaluation whether elements in matrix {\it A} are less-than-or-equal to elements in {\it B}\\
A $>$ ~~~B                   & element-wise evaluation whether elements in matrix {\it A} are greater than elements in {\it B}\\
A $<$ ~~~B                   & element-wise evaluation whether elements in matrix {\it A} are less than elements in {\it B}\\
\hline
\end{tabular}
\end{table}

\begin{table*}[!htb]
\centering
\small
\caption
  {
  Subset of functions for matrix decompositions, factorisations, inverses and equation solvers.
  }
\label{tab:matrix_decompositions}
\vspace{-1ex}
\begin{tabular}{l|lllllllll}
\hline
{\bf Function}   & {\bf Description} \\
\hline
chol(X)          & Cholesky decomposition of symmetric positive-definite matrix {\it X}\\
eig\_sym(X)      & eigen decomposition of a symmetric/hermitian matrix {\it X}\\
eig\_gen(X)      & eigen decomposition of a general (non-symmetric/non-hermitian) square matrix {\it X} \\
\hline
eig\_pair(A, B)   & eigen decomposition for pair of general square matrices {\it A} and {\it B}\\
inv(X)           & inverse of a square matrix {\it X}\\
inv\_sympd(X)     & inverse of symmetric positive definite matrix {\it X}\\
\hline
lu(L, U, P, X)   & lower-upper decomposition of {\it X}, such that {\it PX = LU} and {\it X = P'LU}  \\
null(X)          & orthonormal basis of the null space of matrix {\it X} \\
orth(X)          & orthonormal basis of the range space of matrix {\it X} \\
\hline
pinv(X)          & Moore-Penrose pseudo-inverse of a non-square matrix {\it X}\\
qr(Q, R, X)      & QR decomposition of {\it X}, such that {\it QR = X}\\
qr\_econ(Q, R, X) & economical QR decomposition\\
\hline
qz(AA, BB, Q, Z, A, B) & generalised Schur decomposition for pair of general square matrices {\it A} and {\it B}\\
schur(X)      & Schur decomposition of square matrix {\it X}\\
solve(A, B)       & solve a system of linear equations {\it AX = B}, where {\it X} is unknown\\
\hline
svd(X)           & singular value decomposition of {\it X}\\
svd\_econ(X)     & economical singular value decomposition of {\it X}\\
syl(X)     & Sylvester equation solver\\
\hline
\end{tabular}
\end{table*}

\begin{table}[!htb]
\centering
\small
\caption
  {
  Subset of functions for generating vectors and matrices, showing their main form.
  }
\label{tab:matrix_generators}
\vspace{-1ex}
\begin{tabular}{l|lllllllll}
\hline
{\bf Function}        & {\bf Description} \\
\hline
linspace(start, end, n)     &  vector with {\it n} elements, linearly spaced from {\it start} up to (and including) {\it end} \\
logspace(A, B, n)           &  vector with {\it n} elements, logarithmically spaced from {\it $10^{A}$} up to (and including) {\it $10^{B}$} \\
\hline
regspace(start, $\Delta$, end) &  vector with regularly spaced elements: {\footnotesize [~start,  (start + $\Delta$),  (start + 2$\Delta$),  ...,  (start + $M\Delta$)~ ]},\\
                            &  where {\footnotesize $M$ = floor((end - start)/$\Delta$)}, so that {\footnotesize (start + $M\Delta$) $\leq$ end}\\
\hline
randperm(n, m)              &  vector with {\it m} elements, with random integers sampled without replacement from 0 to {\it n-1} \\
\hline
randu(n\_rows, n\_cols)     &  matrix with random values from a uniform distribution in the [0,1] interval \\
randn(n\_rows, n\_cols)     &  matrix with random values from a normal/Gaussian distribution with $\mu$ = 0 and $\sigma$ = 1\\
\hline
zeros(n\_rows, n\_cols)     &  matrix with all elements set to zero \\
~ones(n\_rows, n\_cols)      &  matrix with all elements set to one \\
\hline
\end{tabular}
\end{table}
\newpage

\begin{table}[!htb]
\centering
\small
\caption
  {
  Subset of general functions of matrices, showing their main form.
  For functions with the {\it dim} argument, {\it dim~=~0} indicates {\it traverse across rows} (ie.~operate on all elements in a~column),
  while {\it dim~=~1} indicates {\it traverse across columns} (ie.~operate on all elements in a~row);
  by default {\it dim~=~0}.
  }
\label{tab:functions_of_matrices}
\vspace{-1ex}
\begin{tabular}{l|lllllllll}
\hline
{\bf Function}        & {\bf Description} \\
\hline
abs(A)                    & obtain element-wise magnitude of each element of matrix {\it A}\\
accu(A)                   & accumulate (sum) all elements of matrix {\it A} into a scalar\\
all(A,dim)                & return a vector indicating whether all elements in each column or row of {\it A} are non-zero\\
\hline
any(A,dim)                & return a vector indicating whether any element in each column or row of {\it A} is non-zero\\
approx{\us}equal(A, B, met, tol)  & return a bool indicating whether all corresponding elements in {\it A} and {\it B} are approx.~equal\\
as{\us}scalar(expression) & evaluate an expression that results in a 1$\times$1 matrix, then convert the result to a pure scalar\\
\hline
clamp(A, min, max)        &  create a copy of matrix {\it A} with each element clamped to be between {\it min} and {\it max}\\
cond(A)                  & condition number of matrix {\it A} (the ratio of the largest singular value to the smallest)\\
conj(C)               & complex conjugate of complex matrix {\it C} \\
\hline
cross(A, B)            & cross product of {\it A} and {\it B}, assuming they are 3 dimensional vectors\\
cumprod(A, dim)             & cumulative product of elements in each column or row of matrix {\it A}\\
cumsum(A, dim)          & cumulative sum of elements in each column or row of matrix {\it A}\\
\hline
det(A)                    & determinant of square matrix {\it A} \\
diagmat(A, k)         & interpret matrix {\it A} as a diagonal matrix (elements not on {\it k}-th diagonal are treated as zero)\\
diagvec(A, k)         & extract the {\it k}-th diagonal from matrix {\it A} (default: {\it k} = 0) \\
\hline
diff(A, k, dim)         & differences between elements in each column or each row of {\it A};~ {\it k} = number of recursions\\
dot(A,B)                  & dot product of {\it A} and {\it B}, assuming they are vectors with equal number of elements \\
eps(A)            & distance of each element of {\it A} to next largest representable floating point number\\
\hline
expmat(A)            & matrix exponential of square matrix {\it A}\\
find(A)               & find indices of non-zero elements of {\it A};~ {\it find(A $>$ k)} finds indices of elements greater than $k$\\
fliplr(A)             & copy {\it A} with the order of the columns reversed\\
\hline
flipud(A)             & copy {\it A} with the order of the rows reversed\\
imag(C)               & extract the imaginary part of complex matrix {\it C}\\
ind2sub(size(A), index) & convert a linear index (or vector of indices) to subscript notation, using the size of matrix {\it A} \\
\hline
inplace{\us}trans(A)     & in-place / in-situ transpose of matrix {\it A} \\
join{\us}rows(A, B)    & append each row of {\it B} to its respective row of {\it A}\\
join{\us}cols(A, B)    & append each column of {\it B} to its respective column of {\it A}\\
\hline
kron(A, B)             & Kronecker tensor product of {\it A} and {\it B}\\
log{\us}det(x, sign, A)   & log determinant of square matrix {\it A}, such that the determinant is exp({\it x})*{\it sign} \\
logmat(A)              & complex matrix logarithm of square matrix {\it A} \\
\hline
min(A, dim)           & find the minimum in each column or row of matrix {\it A}\\
max(A, dim)           & find the maximum in each column or row of matrix {\it A}\\
nonzeros(A)           & return a column vector containing the non-zero values of matrix {\it A} \\
\hline
norm(A,p)             & {\it p}-norm of matrix {\it A}, with {\it p} = 1, 2, $\cdots$, ~ or {\it p} = ``-inf'', ``inf'', ``fro'' \\
normalise(A, p, dim)   & return the normalised version of {\it A}, with each column or row normalised to unit {\it p}-norm \\
prod(A, dim)          & product of elements in each column or row of matrix {\it A}\\
\hline
rank(A)               & rank of matrix {\it A}\\
rcond(A)              & estimate the reciprocal of the condition number of square matrix {\it A} \\
real(C)               & extract the real part of complex matrix {\it C}\\
\hline
repmat(A, p, q)        & replicate matrix {\it A} in a block-like fashion, resulting in {\it p} by {\it q} blocks of matrix {\it A} \\
reshape(A, r, c)      & create matrix with {\it r} rows and {\it c} columns by copying elements from {\it A} column-wise \\
resize(A, r, c)      & create matrix with {\it r} rows and {\it c} columns by copying elements and their layout from {\it A} \\
\hline
shift(A, n, dim)     & copy matrix {\it A} with the elements shifted by {\it n} positions in each column or row \\
shuffle(A, dim)        & copy matrix {\it A} with elements shuffled in each column or row \\
size(A)                & obtain the dimensions of matrix {\it A}\\
\hline
sort(A, direction, dim) & copy {\it A} with elements sorted (in ascending or descending direction) in each column or row \\
sort{\us}index(A, direction)   & generate a vector of indices describing the sorted order of the elements in matrix {\it A}\\
sqrtmat(A)              & complex square root of square matrix {\it A} \\
\hline
sum(A, dim)           & sum of elements in each column or row of matrix {\it A} \\
sub2ind(size(A), row, col)  & convert subscript notation {\it (row,col)} to a linear index, using the size of matrix {\it A} \\
symmatu(A) / symmatl(A) & generate symmetric matrix from square matrix {\it A} \\
\hline
strans(C)            & simple matrix transpose of complex matrix {\it C}, without taking the conjugate\\
trans(A)             & transpose of matrix {\it A} (for complex matrices, conjugate is taken); use {\it A.t()} for shorter form\\
trace(A)                  & sum of the elements on the main diagonal of matrix {\it A}\\
\hline
trapz(A, B, dim)      & trapezoidal integral of {\it B} with respect to spacing in {\it A}, in each column or row of {\it B} \\
trimatu(A) / trimatl(A) & generate triangular matrix from square matrix {\it A} \\
unique(A)             & return the unique elements of {\it A}, sorted in ascending order \\
vectorise(A, dim)          & generate a column or row vector from matrix {\it A} \\
\hline
\end{tabular}
\end{table}

\begin{table}[!htb]
\centering
\small
\caption
  {
  Element-wise functions: matrix {\it B} is produced by applying a function to each element of matrix {\it A}.
  }
\label{tab:elementwise_functions}
\vspace{-1ex}
\begin{tabular}{l|lllllllll}
\hline
{\bf Function}        & {\bf Description} \\
\hline
exp(A) &         base-e exponential: $e^x$ \\ 
exp2(A) &         base-2 exponential: $2^x$ \\ 
exp10(A) &         base-10 exponential: $10^x$ \\
trunc{\us}exp(A) &       base-e exponential, truncated to avoid $\infty$ \\
\hline
log(A) &        natural log: $\log_{e}(x)$ \\
log2(A) &        base-2 log: $\log_{2}(x)$ \\
log10(A) &        base-10 log: $\log_{10}(x)$ \\
trunc{\us}log(A) &       natural log, truncated to avoid $\pm\infty$ \\ 
\hline
pow(A, p) &        raise to the power of p: $x^p$ \\
square(A) &        square: $x^2$ \\
sqrt(A) &        square root: $\sqrt x$ \\
\hline
floor(A) &       largest integral value that is not greater than the input value \\
ceil(A) &       smallest integral value that is not less than the input value \\
round(A) &       round to nearest integer, with halfway cases rounded away from zero \\
trunc(A) &       round to nearest integer, towards zero \\
\hline
erf(A) &       error function \\
erfc(A) &       complementary error function \\
lgamma(A) &       natural log of the gamma function \\
\hline
sign(A) &       signum function; for each element {\it a} in {\it A}, the corresponding element {\it b} in B is: \\
        &  $b = \left\{ \begin{array}{ccc}
-1 & \mbox{if} & a < 0 \\
~0 & \mbox{if} & a = 0 \\
+1 & \mbox{if} & a > 0 \end{array} \right.  $ \\
\hline
trig(A) & trignometric function, where {\it trig} is one of:\\
        & {\it cos}, {\it acos}, {\it cosh}, {\it acosh}, {\it sin}, {\it asin}, {\it sinh}, {\it asinh}, {\it tan}, {\it atan}, {\it tanh}, {\it atanh} \\
\hline
\end{tabular}
\end{table}

\begin{table}[!htb]
\centering
\small
\caption
  {
  Subset of functions for statistics, showing their main form.
  For functions with the {\it dim} argument, {\it dim~=~0} indicates {\it traverse across rows} (ie.~operate on all elements in a~column),
  while {\it dim~=~1} indicates {\it traverse across columns} (ie.~operate on all elements in a~row);
  by default {\it dim~=~0}.
  }
\label{tab:statistics_functions}
\vspace{-1ex}
\begin{tabular}{l|lllllllll}
\hline
{\bf Function/Class}        & {\bf Description} \\
\hline
cor(A, B)              & generate matrix of correlation coefficients between variables in {\it A} and {\it B}\\
cov(A, B)              & generate matrix of covariances between variables in {\it A} and {\it B}\\
\hline
hist(A, centers, dim) & generate matrix of histogram counts for each column or row of {\it A}, using given bin centers\\
histc(A, edges, dim) & generate matrix of histogram counts for each column or row of {\it A}, using given bin edges\\
kmeans(means, A, k, ...) &  cluster column vectors in matrix {\it A} into {\it k} disjoint sets, storing the set centers in {\it means}\\
\hline
princomp(A)      & principal component analysis of matrix {\it A}\\
running{\us}stat     & class for running statistics of a continuously sampled one dimensional signal \\
running{\us}stat{\us}vec     & class for running statistics of a continuously sampled multi-dimensional signal\\
\hline
mean(A, dim)        & find the mean in each column or row of matrix {\it A} \\
median(A, dim)      & find the median in each column or row of matrix {\it A}\\
stddev(A, norm{\us}type, dim)      & find the standard deviation in each column or row of {\it A}, using specified normalisation\\
var(A, norm{\us}type, dim)         & find the variance in each column or row of matrix {\it A}, using specified normalisation\\
\hline
\end{tabular}
\end{table}
\newpage

\begin{table}[!htb]
\centering
\small
\caption
  {
  Subset of functions for signal and image processing, showing their main form.
  }
\label{tab:signal_processing_functions}
\vspace{-1ex}
\begin{tabular}{l|lllllllll}
\hline
{\bf Function/Class}        & {\bf Description} \\
\hline
conv(A, B)             & 1D convolution of vectors {\it A} and {\it B}\\
conv2(A, B)             & 2D convolution of matrices {\it A} and {\it B}\\
\hline
fft(A, n)             & fast Fourier transform of vector {\it A}, with transform length {\it n}\\
ifft(C, n)             & inverse fast Fourier transform of complex vector {\it C}, with transform length {\it n}\\
\hline
fft2(A, rows, cols)             & fast Fourier transform of matrix {\it A}, with transform size of {\it rows} and {\it cols}\\
ifft2(C, rows, cols)             & inverse fast Fourier transform of complex matrix {\it C}, with transform size of {\it rows} and {\it cols}\\
\hline
interp1(X, Y, XI, YI) & given a 1D function specified in vectors {\it X} (locations) and {\it Y} (values), \\
                      & generate vector {\it YI} containing interpolated values at given locations {\it XI}\\
\hline
interp2(X, Y, Z, XI, YI, ZI) & given a 2D function specified by matrix {\it Z} with coordinates given by vectors {\it X} and {\it Y}, \\
                             & generate matrix {\it ZI} containing interpolated values at coordinates given by vectors {\it XI} and {\it YI} \\
\hline
\end{tabular}
\end{table}

\begin{table}[!htb]
\centering
\small
\caption
  {
  Examples of Matlab/Octave syntax and conceptually corresponding PyArmadillo syntax.
  Note that for submatrix access the exact conversion from Matlab/Octave to PyArmadillo syntax
  will require taking into account that indexing starts at 0.
  }
\label{tab:matlab_armadillo_syntax}
\vspace{-1ex}
\begin{tabular}{l|l|lllll}
\hline
{\bf Matlab \& Octave} & {\bf PyArmadillo} & {\bf Notes} \\
\hline
A(1, 1)             & A$[$0, 0$]$      & indexing in PyArmadillo starts at 0, following Python convention \\
A(k, k)             & A$[$k-1, k-1$]$  & \\
\hline
size(A,1)           & A.n\_rows        & member variables are read only \\
size(A,2)           & A.n\_cols        & \\
size(Q,3)           & Q.n\_slices      & Q is a cube (3D array) \\
numel(A)            & A.n\_elem        & .n\_elem indicates the total number of elements \\
\hline
A(:, k)             & A$[$:, k$]$      & read/write access to a specific column \\
A(k, :)             & A$[$k, :$]$      & read/write access to a specific row    \\ 
A(:, p:q)           & A$[$:, p:q$]$    & read/write access to a submatrix spanning the specified cols \\
A(p:q, :)           & A$[$p:q, :$]$    & read/write access to a submatrix spanning the specified rows \\
A(p:q, r:s)         & A$[$p:q, r:s$]$  &  \\
\hline
Q(p:q, r:s, t:u)    & Q$[$p:q, r:s, t:u$]$ & Q is a cube (3D array) \\
Q(:, :, k)          & Q$[$:, :, k$]$  & access one slice\\
Q(:, :, k)          & Q$[$single\_slice, k$]$  & access one slice, with the slice provided as a matrix \\
\hline
A\hspace{1pt}$'$    & A\hspace{0.5pt}.\hspace{0.5pt}t() ~or~ trans(A)   & transpose (for complex matrices the conjugate is taken) \\ 
A\hspace{1pt}$.'$   & A\hspace{0.5pt}.\hspace{0.5pt}st() ~or~ strans(A) & simple transpose (for complex matrices the conjugate is not taken) \\ \hline
A = zeros(size(A))  & A.zeros()             & set all elements to zero \\
A = ones(size(A))   & A.ones()              & set all elements to one \\ 
A = zeros(k)        & A = zeros(k, k) & create a matrix with elements set to zero \\       
A = ones(k)         & A = ones(k, k)  & create a matrix with elements set to one  \\ \hline
\multirow{2}{*}{C = complex(A,B)}    & \multirow{2}{*}{C = cx\_mat(A,B)} & \multirow{2}{*}{construct a complex matrix from two real matrices} \\
                    &                          & \\ \hline
A $*$ ~B            & A $*$ ~B                  & {\footnotesize$*$} indicates matrix multiplication \\
A $.*$ B            & A {\footnotesize$@$} B                  & {\footnotesize$@$} indicates element-wise multiplication \\
A $./$ B            & A $/$ B                   & $/$ ~indicates element-wise division \\
A $\backslash$ ~B    & solve(A,B)              & solve a system of linear equations \\
A = A $+$ 1        & A $+$= 1                   & \\
A = A $-$ 1        & A $-$= 1                 & \\ \hline
A = [ 1 ~ 2;       & A = mat($[$ $[$ 1, 2 $]$,     & \\
~~~~~ ~ ~ 3 ~ 4; ] & ~~~~~~~~~~~~~~~~~~$[$ 3, 4 $]$  $]$)  & \\
\hline
X = [ A~ B ]       & X = join\_rows(A,B)     & \\
X = [ A; B ]       & X = join\_cols(A,B)     & \\ \hline
A & A.print(``A:'')  &  print the entire contents of A \\
save -ascii `A.txt' A &  A.save(``A.txt'', raw\_ascii) &  Matlab/Octave matrices saved as ascii text are readable\\
load -ascii `A.txt'   &  A.load(``A.txt'', raw\_ascii) &  by PyArmadillo (and vice-versa) \\ \hline
A = rand(2,3)      & A = randu(2,3)   &  {\it rand\underline{u}} generates uniformly distributed random numbers\\
B = randn(4,5)     & B = randn(4,5)   &  \\
\hline
\end{tabular}
\end{table}

\clearpage
\section*{Acknowledgements}

We would like to thank our colleagues at Data61/CSIRO
(Dan Pagendam, Dan Gladish, Andrew Bolt, Piotr Szul)
for providing feedback and testing.

\small
\bibliographystyle{ieee}
\bibliography{references}

\end{document}